\documentclass[conference,a4paper]{IEEEtran}

\usepackage{graphicx, caption, subcaption}
\usepackage{algorithm}
\usepackage{array}

\usepackage{algorithmicx}
\usepackage{algpseudocode}

\usepackage{soul}
\usepackage{url}
\usepackage{verbatim}
\usepackage{cite}
\usepackage{listings}
\usepackage[table]{xcolor}
\usepackage[nolist]{acronym}
\usepackage{amsmath}
\usepackage{amssymb}
\usepackage{soul}
\usepackage{rotating}
\usepackage{multirow} 

\usepackage{mathtools, stmaryrd}
\usepackage{xparse} \DeclarePairedDelimiterX{\Iintv}[1]{\llbracket}{\rrbracket}{\iintvargs{#1}}
\NewDocumentCommand{\iintvargs}{>{\SplitArgument{1}{,}}m}
{\iintvargsaux#1} %
\NewDocumentCommand{\iintvargsaux}{mm} {#1\mkern1.5mu..\mkern1.5mu#2}

\usepackage{siunitx}

\usepackage{blindtext}
\usepackage{ctable} 
\def\BibTeX{{\rm B\kern-.05em{\sc i\kern-.025em b}\kern-.08em
    T\kern-.1667em\lower.7ex\hbox{E}\kern-.125emX}}

\begin{document}

\voffset=0.05in
\textheight=9.28in

\title {Efficient Self-Learning and Model Versioning for AI-native O-RAN Edge}
\author{\IEEEauthorblockN{
        Mounir Bensalem\IEEEauthorrefmark{1}, 
        Fin Gentzen\IEEEauthorrefmark{1}, 
        Tuck-Wai Choong\IEEEauthorrefmark{2}, 
        Yu-Chiao Jhuang\IEEEauthorrefmark{2}, 
        Admela Jukan\IEEEauthorrefmark{1},  
        Jenq-Shiou Leu\IEEEauthorrefmark{2}}
\IEEEauthorblockA{
    \IEEEauthorrefmark{1}Technische Universit\"at Braunschweig, Germany;\\
\{mounir.bensalem, f.gentzen, a.jukan\}@tu-bs.de\\
 \IEEEauthorrefmark{2}National Taiwan University of Science and Technology, Taipei, Taiwan\\ \{d11302001, d10902011, jsleu\}@mail.ntust.edu.tw}
}

\maketitle

\begin{abstract}  
The AI-native vision of 6G requires Radio Access Networks to train, deploy, and continuously refine thousands of machine learning (ML) models that drive real-time radio network optimization. Although the Open RAN (O-RAN) architecture provides open interfaces and an intelligent control plane, it leaves the life-cycle management of these models unspecified. Consequently, operators still rely on ad-hoc, manual update practices that can neither scale across the heterogeneous, multi-layer stack of Cell-Site, Edge-, Regional-, and Central-Cloud domains, nor across the three O-RAN control loops (real-, near-real-, and non-real-time).  
We present a self-learning framework that provides an efficient closed-loop version management for an AI-native O-RAN edge. In this framework, training pipelines in the Central/Regional Cloud continuously generate new models, which are cataloged along with their resource footprints, security scores, and accuracy metrics in a shared version repository. An Update Manager consults this repository and applies a self-learning policy to decide \emph{when} and \emph{where} each new model version should be promoted into operation. A container orchestrator then realizes these decisions across heterogeneous worker nodes, enabling multiple services (rApps, xApps, and dApps) to obtain improved inference with minimal disruption.  
Simulation results show that an efficient RL-driven decision-making can guarantee quality of service, bounded latencies while balancing model accuracy, system stability, and resilience.

\end{abstract}



\section{Introduction}
AI-Native integration has emerged as a core feature of 6G, with artificial intelligence (AI) distributed across all network layers to support new services and enable network automation and optimization \cite{wu2022ai}. The Open Radio Access Network (O-RAN) architecture, founded on openness, intelligence, virtualization, and interoperability \cite{oran_arch_2025}, offers the programmable control plane (SMO/RIC) and standardized interfaces required to embed AI capabilities.  However, realizing an \emph{autonomous} and \emph{self-learning} 6G O-RAN edge presents an important operational challenge: how to manage the life-cycle of a rapidly growing amount of ML models that must be re-trained, versioned, and rolled out at scale while guaranteeing service continuity. On the other hand, towards achieving AI-native vision of O-RAN, the ability to train, deploy, and continuously refine machine-learning (ML) models is critical for radio-resource management, traffic steering, energy savings, security enforcement, and a host of emerging 6G services. 

In radio environments, traffic patterns and user behaviors can change in minutes.  To stay effective, RAN-optimization models (e.g., those driving beamforming or hand-over decisions) require frequent re-training on fresh data.  Each training event yields a new model version whose benefits (higher accuracy) must be weighed against potential hazards (instability, security vulnerabilities, resource spikes).  In today’s deployments, operators rely on manual or heuristic update policies that do not scale to the hundreds of rApps/xApps/dApps foreseen for 6G, and the O-RAN specifications intentionally leave life-cycle management to vendor-specific solutions.  Consequently, networks risk either lagging behind the state-of-the-art or suffering disruptive roll-outs. This is an open issues in today's networks that needs to be urgently addressed.

This paper proposes for the first time an efficient, self-learning ML model versioning  management for an AI-native O-RAN edge.  We design the first end-to-end ML life-cycle framework that unifies cloud-based training with multi-layer O-RAN inference, explicitly accounting for accuracy, reliability, security, and stability across the three O-RAN control loops.  We propose a practical RL-driven update policy that leverages fine-grained telemetry and a version repository to maximize network utility while safeguarding stability. Simulation results show that in a dynamic environment when AI-based RIC applications (rApps/xApps/dApps) might requires updates, RL tends to favor stability over accuracy improvement of ML models for dApps while preserving delay constraints. At the same time, the framework tends to update ML models used by xApps and rApps for the scenarios where updating ML models has minimal impact on delay constraints. 
\par The rest of the paper is organized as follows. Section \ref{sec:related} presents the related work. 
Section \ref{sec:model} describes the reference architecture. Section \ref{sec: problem} formulates the ML model versioning optimization problem and the RL-based solution.  
Section \ref{sec:results} presents numerical results. Section \ref{sec:conclusion} concludes the paper.

\begin{figure*}
 \centering 
    \includegraphics[scale=0.80]{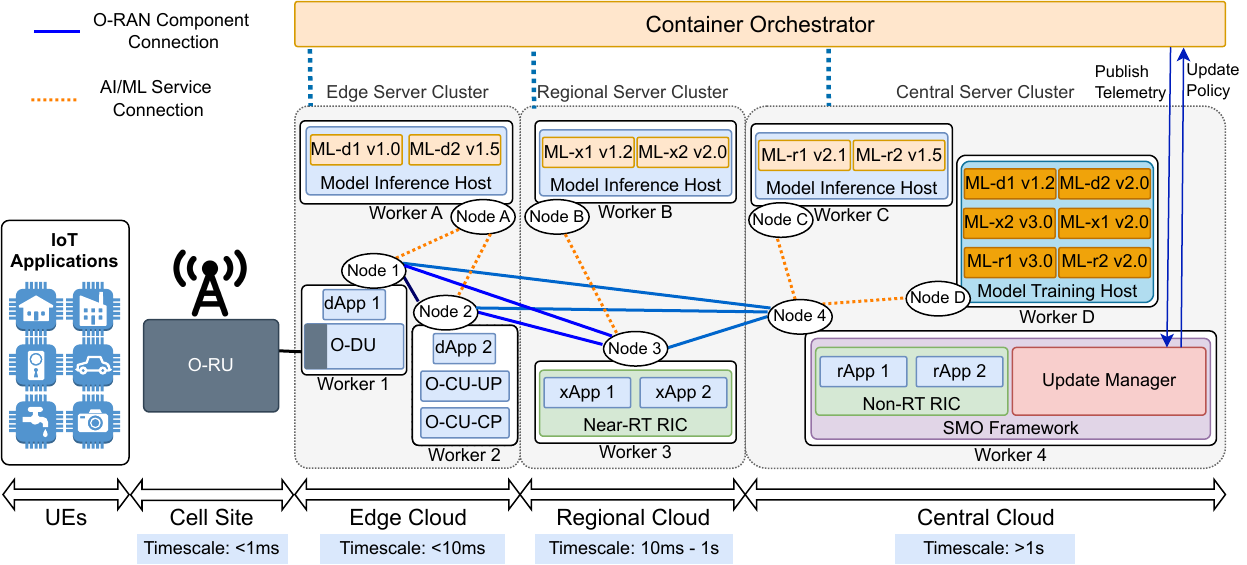}
 \caption{O-RAN compliant reference architecture with self-learning and ML model versioning.}
\label{fig:system}
\end{figure*}

\section{Related Work}\label{sec:related}

The past decade has witnessed an explosive growth in the size and complexity of ML models, accompanied by continual improvements in accuracy and robustness \cite{paleyes2022challenges}.  To remain effective, deployed models must be re-trained and re-deployed whenever data distributions, operating conditions, or security requirements change \cite{angioni2024robustness}.  This calls for rigorous versioning of  models, data, and code, a problem recognized by the MLOps community as one of its grand challenges \cite{kreuzberger2023machine}.  

Early studies concentrated on updating a single model in cloud settings, e.g., forecasting data-series in \cite{10597011}.  More recent work has shifted toward edge deployments, where limited resources and latency constraints exacerbate the trade-off between accuracy, reliability, security, and system stability \cite{45305}.  Prior research—including our own optimization framework for edge-only scenarios \cite{gentzen2024effective}, has shown that reinforcement‐learning (RL) agents can automate version selection by balancing accuracy, stability, and security.  However, those studies assume a homogeneous edge setting and treat model training as an external, offline process.  The hierarchical O-RAN stack that spans Cell-Site, Edge-, Regional-, and Central-Cloud layers was not studied, in which there is a need to coordinate updates across treal-time, near-real-time, non-real-time control loops.

O-RAN seeks to disaggregate the classical base station into an Open Radio Unit (O-RU), Open Distributed Unit (O-DU), and Open Central Unit (O-CU), interconnected through standardized open interfaces such as the Open Fronthaul and F1/Mid-haul \cite{oran_arch_2025}.  Additional management and intelligence are provided by the Service Management and Orchestration (SMO) framework and by two RAN Intelligent Controllers: the Near-Real-Time RIC (running \emph{xApps}, $10~\mathrm{ms}$–$1~\mathrm{s}$ loops) and the Non-Real-Time RIC (running \emph{rApps}, loops $>1~\mathrm{s}$).  Very recent work further proposes \emph{dApps} inside the CU/DU to provide sub-$10~\mathrm{ms}$ decision making \cite{oran_dapp}.  

Existing MLOps research does not consider the demanding latency, heterogeneous hardware, and layered control loops of O-RAN, whereas O-RAN-specific studies do not study, systematic, data-driven version management.  To the best of our knowledge, there is still no \emph{end-to-end} framework that (i) integrates cloud-based training with edge/near-RT inference, (ii) accounts for multi-objective trade-offs (accuracy, reliability, security, and stability), and (iii) scales across the Cell-Site, Edge-, Regional-, and Central-Cloud layers while respecting the real-time, near-real-time, and non-real-time control loops. This paper fills this knowledge gap by introducing a self-learning framework for \emph{large-scale model version management in AI-native 6G O-RAN}.  Our solution embraces the full O-RAN stack, coordinates updates across all three control loops, and is the first to unify cloud-side training with distributed inference while explicitly optimizing for accuracy, security, reliability, and service continuity.

\section{Reference Architecture}\label{sec:model}

Figure~\ref{fig:system} presents the reference architecture adopted in this work, which follows the multi-layer structure  and key components compliant with the O-RAN specifications~\cite{oran_cloud_arch}. The system is structured into multiple functional context layers, starting from User Equipment (UE) over to the central cloud. At the \emph{Cell-Site} layer, O-RUs carry out RF processing; the cloudification of these units and the definition of an O-RU Accelerator Abstraction Layer (AAL) are still open research topics \cite{oran_understanding}. The \emph{Edge-Cloud} layer hosts the O-DU and O-CU functions. The \emph{Regional-Cloud} layer accommodates the Near-RT RIC, and (iv) the \emph{Central-Cloud} layer contains the Non-RT RIC together with SMO services. A salient feature of these layered architecture is in its concentric control loops~\cite{oran_understanding}: (i) a non-real-time loop (periods $>1$s) executed by rApps in the Non-RT RIC for policy management and long-term optimization; (ii) a near-real-time loop (10ms - 1s) handled by xApps in the Near-RT RIC for short- to medium-term resource control; and, (iii) a real-time loop ($<10$ms) implemented by dApps embedded within the E2-connected CU/DU nodes for immediate radio-access decisions~\cite{oran_dapp}. Together, these layers and feedback loops provide the hierarchical control framework that supports the proposed networking solution.

In the depicted architecture, the ML models reside on O-Cloud compute nodes, while dApps, xApps, and rApps obtain inference by invoking the Model Inference Host running on Workers A–C. Figure \ref{fig:system} further illustrates the ML model version management pipeline. Training jobs running in Worker D (e.g., ML-d1-v1.2 representing an ML task used by dApps, and ML-x2 v2.0 for xApps) generate candidate models; after validation, specific versions (e.g., ML-d1 v1.0, ML-x1 v1.2) are promoted to production.
Additionally, Figure \ref{fig:system} shows the Update Manager as an independent component responsible for coordinating ML model update strategies and releases. The Container Orchestrator is responsible for the actual deployment and management of ML model containers, ensuring smooth model updates. This design allows the system to flexibly deploy different versions of ML models on different computing resources (such as Workers A-D), thereby achieving efficient version management and resource utilization.

\subsection{ML Update Manager}

\begin{figure}
 \centering 
    \includegraphics[scale=0.52]{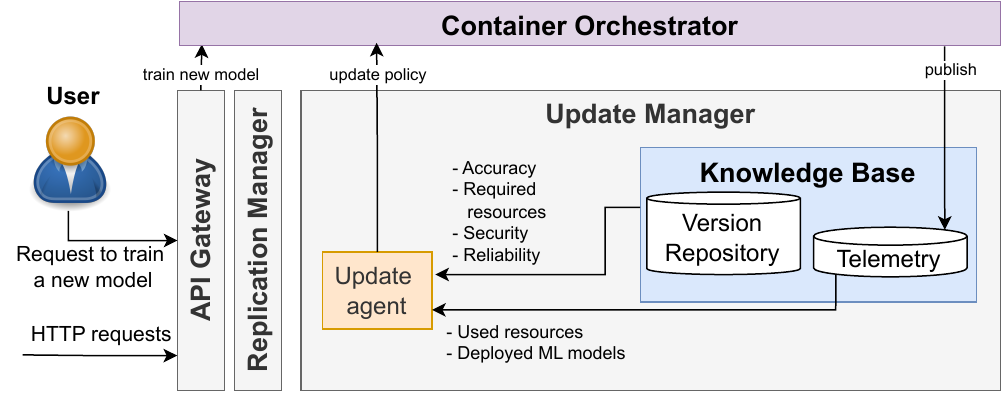}
 \caption{ML Update Manager }
\label{fig:solution}
\end{figure}

The \textit{ML Update Manager} uses data from the Telemetry, Knowledge Base, and Update Agents. Telemetry continuously captures operational data, including notifications of newly available ML model versions pushed from the cloud, and stores it for later analysis. Knowledge Base is assumed to hosts ML Version Repository that stores every model release along with its resource footprint and quality attributes, such as accuracy, stability, reliability, and security scores. Update Agent uses runs algorithms to decide whether an update of the ML model should be executed immediately, or delayed.

The ML Update Manager decides whether a running ML-inference replica should keep its current ML model or switch to a newly released one. After every inference request, the Update Agent checks if a fresher model version is available. If it finds one, it weighs the merits of leaving the replica untouched versus terminating it and spawning a new instance that uses the latest model. The logic behind this choice may be based on anything from simple domain-specific heuristics to advanced optimization routines or reinforcement-learning agents trained to discover near-optimal policies. 

For lifecycle management, each model release carries two identifiers: a major version X and a minor version Y, expressed as X.Y. A major upgrade (X) signals substantial changes, e.g., a new network architecture or significant feature additions—and is usually rolled out at predetermined intervals to avoid service disruption. Minor updates (Y) deliver targeted improvements, such as security patches or reliability fixes, and can be issued at any time in response to alerts, user feedback, or live performance metrics. 

\subsection{A Sample ML-based workflow in O-RAN}

\begin{figure}[t]
 \centering 
   \includegraphics[scale=0.6]{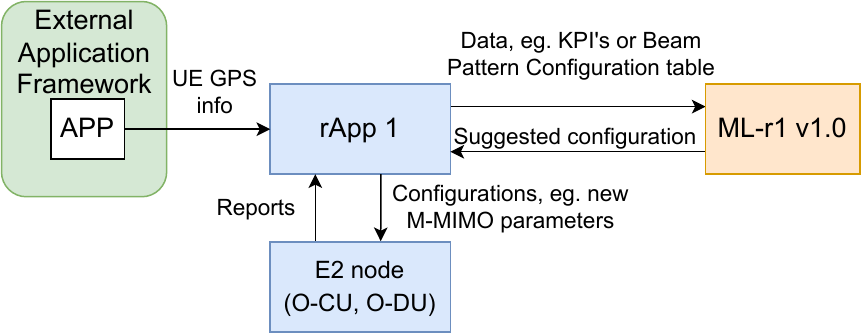}
 \caption{A simplified ML versioning flow diagram for massive MIMO beamforming optimization, based on \cite{oran_usecases_2025} }\label{subfig:f1}
\end{figure}

Building on the O-RAN Alliance use-case catalogue  \cite{oran_usecases_2025}, which documents multiple scenarios that embed AI/ML inference and training in the RAN, we illustrate the ML workflow in Figure \ref{subfig:f1}. Here, the example is provided for Massive-MIMO beamforming, based on Figure 4.6.3.1-1 from \cite{oran_usecases_2025}. 

In general, External Application Framework supplies Enrichment Data to RIC applications; rApps, xApps, and dApps. These applications fuse the enrichment data inputs with live network measurements, execute ML-enabled models, and produce inference results that are translated into control Actions and Reports exchanged with E2-connected nodes (O-CU, O-DU). The resulting closed loop drives continuous optimization of radio-network performance. The ML models are denoted by ML-ID with a version (x.y) that can be periodically updated  based on measurements data to provide accurate predictions and recommendations that improve the performance of the system. Here,  the Application Framework forwards UE GPS coordinates alongside key performance indicators (KPIs) and current beam-pattern settings to rApp 1. The model \textit{ML-r1 v1.0} analyses this data and outputs recommended beam-configuration updates. These recommendations are applied by E2 nodes, which subsequently return updated performance metrics to rApp 1, in an ongoing, self-optimizing cycle.

\section{ML Model Versioning with RL} \label{sec: problem}
\subsection{Assumptions}
We consider a system deployment, including a set of model inference host nodes $\epsilon = \{ E_1, \ldots, E_n, \ldots, E_N \}$, a set of O-RAN nodes running RIC applications in the edge, regional and central cloud, $\mathcal{R} = \{ R^e,  R^r, \ldots, R^c \}$, where  $ R^e = \{ R^e_1, \ldots, R^e_{n}, \ldots, R^e_{N_e} \}$ are the nodes in the edge cloud, $ R^r = \{ R^r_1, \ldots, R^r_{n}, \ldots, R^r_{N_r} \}$  the nodes in the regional cloud, and $ R^c = \{ R^c_1, \ldots, R^c_{n}, \ldots, R^c_{N_c} \}$  the nodes in the central cloud.
We assume that each node is constrained by a certain amount of capacity modeled
as a number of CPU units $C_n$, RAM space $S_n$, and Disk space $D_{n}$.
 We consider a set of ML models denoted as $\kappa = \{ 1, \ldots, k, \ldots, K \}$. We assume that each ML model $k$, can have a  version $x\in [0, X]$, 
  We define a decision variable $a=1$ when updating a ML model $k$ of version $x$ to a newer version $x+1$, 0 otherwise. We define a mapping function $g(x,a)$ that gives the new version of an ML model $k$ of version $x$ after a decision update $a$. Each version $x$ of model $k$ requires resources $b_{k,x} = [b_{k,x}^\text{cpu}, b_{k,x}^\text{ram}, b_{k,x}^\text{disk}]$, 
of CPU units, RAM, and disk space.

We denote by $\delta_{k,x}(n)$ the number of replicas from ML model $k$ of version $x$ allocated in node $E_n$. The resource allocation has
the capacity constraints, i.e.,

\begin{equation}\label{eq:cpus}
    \sum_{k=1}^{K} \sum_{x=1}^{X}  b_{k,x}^\text{cpu} \delta_{k,x}(n) \leq C_n, \quad \forall n \in [1, N]
\end{equation}
\begin{equation}\label{eq:rams}
    \sum_{k=1}^{K}\sum_{x=1}^{X} b_{k,x}^\text{ram} \delta_{k,x}(n) \leq R_n, \quad \forall n \in [1, N]
\end{equation}
\begin{equation}\label{eq:disks}
    \sum_{k=1}^{K}\sum_{x=1}^{X}  b_{k,x}^\text{disk} \delta_{k,x}(n) \leq D_n, \quad \forall n \in [1, N]
\end{equation}

The equations (\ref{eq:cpus})-(\ref{eq:disks}) assures that the replicas created in a node $E_n$ does not require more that its CPU, RAM and Disk capacity. We consider 5 types of delays: inference delay, processing delay (RIC applications), transmission delay, delay to spawn a replica (where applicable) and queuing delay. 
The processing delay $\tau^p_{k}$ of a RIC application using an ML model
$k$ is including all O-RAN operations. The inference delay $\tau^I_{k,x,n}$ of an ML model
$k$ of version $x$ deployed through a pod in node $E_n$ is distributed around an average value. We consider transmission and propagation times to be a known parameter since they depend on the distance between model inference host nodes and the nodes that running RIC applications.
The routing and path computation is typically managed by the container orchestrator (e.g., Kubernetes). 
We use $F$ to denote the total number of requests to a model $k$. The time to spawn a new replica, $\tau^s_{k,n}(a)$, is assumed constant and occurs when resources are scaled or when the Update Manager replaces version $x$ with  $x+1$. If no update is performed ($a=0$), the spawn time is zero.
We assume the system has a queuing buffer for each ML model $k$. We denote by $\tau^q_{f(k)}$  the total queuing delay of a request  $f(k)$ of  ML model $k$.  
The queuing delay is measured as the difference between  the departure request time and the arrival request time plus the transmission plus the spawn time and the processing and inference time. Finally, the total delay $\tau_{f(k)}$ for each  request $f(k)$  of ML model $k$, processed using a replica that runs version $x$ is given as:
\begin{equation}\begin{split}
    \tau_{f(k)}(x, a) =& \tau^p_{k} + \tau^I_{k,g(x,a),n}+ \tau^t_{k,n} + \tau^s_{k,n}(a)  
    \\ &+ \tau^q_{f(k)}, \quad  \forall k \in [1, K]
    \end{split}
\end{equation}
The average  delay for all requests, is defined as, i.e., 
\begin{equation}
    \mathcal{O}_1(\Delta, a) = \frac{1}{KF} \sum_{k=1}^{K}\sum_{f=1}^{F} \tau_{f(k)} (x,y, a)
\end{equation}
where $\Delta$ describes the state of the network, including nodes, and deployed replicas of ML models and their versions.

After training an ML model $k$, we obtain a new version that is assumed to have a higher accuracy.  We denote by  $Acc_{k}$($x, a$) the accuracy of a ML model $k$ with version $x$ after taking an update decision $a$. The average accuracy of  all requests is given by:
\begin{equation}
    \mathcal{O}_2(\Delta, a) = \frac{1}{KF} \sum_{k=1}^{K}\sum_{f=1}^{F} Acc_{k}(x, a)
\end{equation}

We assume that updates can improve the stability of an ML model operation, defined as parameters $St_{k}(x, a)$, assigned to ML model $k$ with version $x$ after taking an update decision $a$. We finally define an objective function to yield an average among all requests as solution, i.e.,
\begin{equation}
    \mathcal{O}_3(\Delta, a) = \frac{1}{KF} \sum_{k=1}^{K}\sum_{f=1}^{F} St_{k}(x, a)
\end{equation}

\subsection{Problem Formulation}
We define our ML Model Versioning Problem (MMV)  as a multi-objective optimization problem, aiming at maximizing the average ML model accuracy, stability of all processed requests while minimizing the average delay. The MMV problem is formulated as follows:
\begin{equation}\begin{split}
    \max_{a\in \mathcal{A} } & \left( -  \mathcal{O}_1(\Delta, a), \mathcal{O}_2(\Delta, a), \mathcal{O}_3(\Delta, a),  \right)  \\
    \text{subject to:   }& {Eq.} (\ref{eq:cpus}), (\ref{eq:rams}), (\ref{eq:disks})
\end{split}
\end{equation}

\begin{algorithm}
\footnotesize
    \caption{RL-based Model Versioning}
    \label{alg:UpdateDecision}
    \begin{algorithmic}[1]
        \State \textbf{Input:} Event list; $b_k$, 
        Network state $(Z, Q)$
        \State \textbf{Initialization:} Q-table $\mathcal{Q}_t$, exploration rate $\epsilon$, learning rate $\alpha_{lr}$, discount factor $\gamma$, max episodes, $\epsilon$-decay
        \ForAll {events}
        \If{an update is possible or scaling is triggered}
            \State $s \leftarrow (Z, Q, e, f)$
            \State $a \leftarrow \textsc{UpdateDecision}(s)$  \Comment{chooses action agnostic of $\epsilon$-greedy}
            \State $Z', Q', e', f' \leftarrow \text{GetNextState}(\cdot)$
            \State $s' \leftarrow (Z', Q', e', f')$
            \State $\mathcal{Q}_t \leftarrow \textsc{UpdateQTable}(s, a, s')$
            \State $s \leftarrow s'$
        \Else
            \State \textbf{pass}
        \EndIf
        \If{$\epsilon > \epsilon_{\min}$}
            \State $\epsilon \leftarrow \epsilon \cdot \text{decay}$
        \EndIf
        \EndFor
    \end{algorithmic}
\end{algorithm}

\subsection{RL-Based Update Decision}\label{sec:solution}
Our RL solution comprises the usual components: \emph{agent}, \emph{state space} $S$, \emph{action space} $A$, \emph{reward} $R$, and the \emph{environment}. The agent observes metrics such as delay, security, reliability, accuracy, and stability for each update decision and uses a Q-table to track state/action pairs and the resultant rewards.

State space is denoted by $S = \{\, s \mid s = (Z_k, f_k, Q_k, e, \Omega) \}$, where \emph{i)} $Z_k = \{\zeta_{1,k}, \ldots, \zeta_{n,k}, \ldots, \zeta_{N,k}\}$ captures the capacity status of each node $n$ for model $k$ (CPU, RAM, and Disk).  
\emph{ii)} $f_k$ is the integer identifier for the ML model in use.  
\emph{iii)} $Q_k$ is an integer variable representing the queue size for model $k$.

 The reward $R$ combines delay $\psi$, accuracy $\upsilon$, and stability $\sigma$, representing the disruption introduced by frequent updates:
\begin{equation}
    R(s,a) \;=\; -(1-\alpha)(w_1\,\psi \;+\; w_2\,\sigma) \;+\; \alpha w_3\,\upsilon,
    \label{eq:reward}
\end{equation}
where higher values of $\vartheta$, $\eta$, and $\upsilon$ are desired, whereas delay $\psi$ and destabilization $\sigma$ are penalized. The weighting factors $w_i \ge 0$ can be tuned to emphasize different objectives, while $\alpha$ is coefficient that controls  the trade-off between the  maximization and minimization of different goals.

Algorithm~\ref{alg:UpdateDecision} illustrates our self-learning procedure for update decisions. At each event (arrival or departure), the agent observes the system state $s$, consults its Q-table to select an action $a$ (balancing exploration and exploitation via $\epsilon$-greedy), and computes the subsequent state $s'$. The Q-table is then updated accordingly. Over repeated episodes, the agent converges toward policies that balance optimal model updates in terms of delay, accuracy, security, reliability, and stability.

In the early training phase, the agent executes \emph{exploratory} actions with probability $\epsilon$ to discover a diverse set of state-action paths. As the agent gathers more experience, $\epsilon$ diminishes until reaching a minimal exploration rate, favoring \emph{exploitation} of learned policies.
 Each node $n$ has CPU, RAM, and disk capacities, as well as an associated transmission times. For instance, higher latencies for the cloud node can also be integrated as part of $Z$, $Q$, and $e$ within the RL environment. By applying the RL-based approach, the system adaptively learns whether updating at a given node configuration is beneficial or whether continuing with an existing version is more favorable, considering performance, resource constraints, and stability factors.  


\section{Numerical Evaluation}\label{sec:results}

\begin{figure*}
\centering
    \begin{subfigure}[t]{0.32\textwidth}
    \centering
        \includegraphics[scale=0.23]{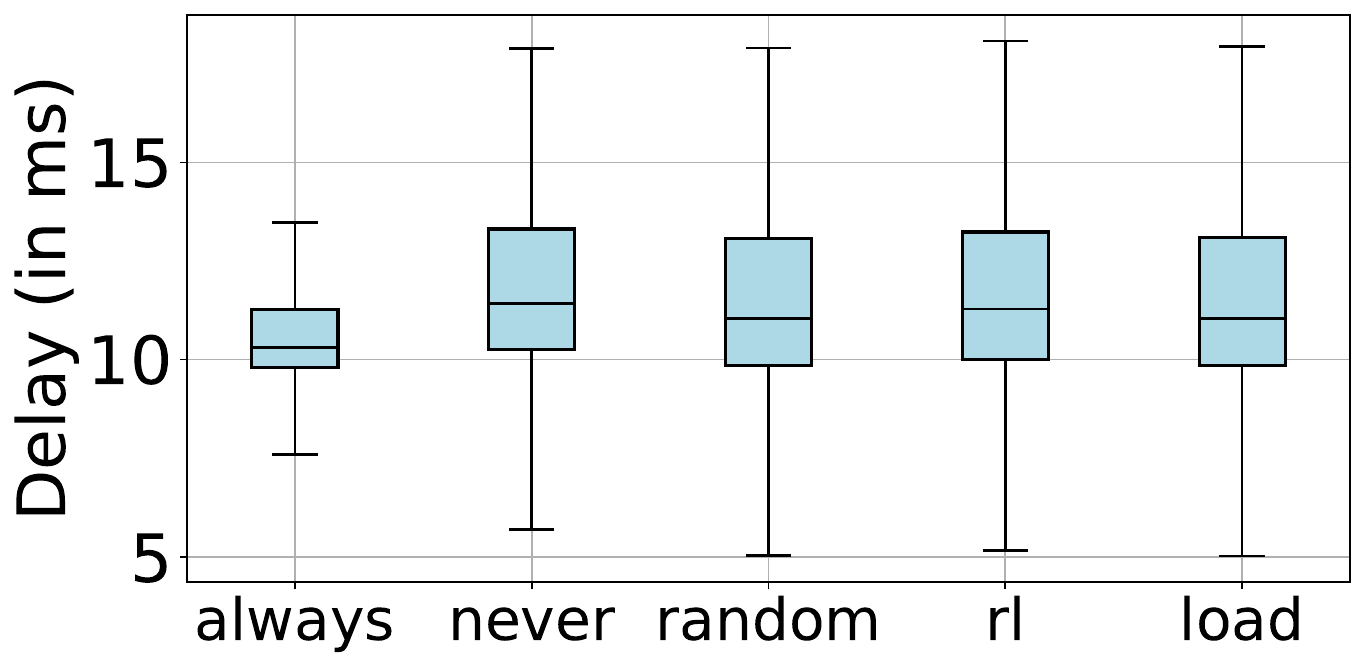}
        \caption{ORAN-dApp}
        \label{subfig:d1}
    \end{subfigure}
    \begin{subfigure}[t]{0.32\textwidth}
    \centering
        \includegraphics[scale=0.23]{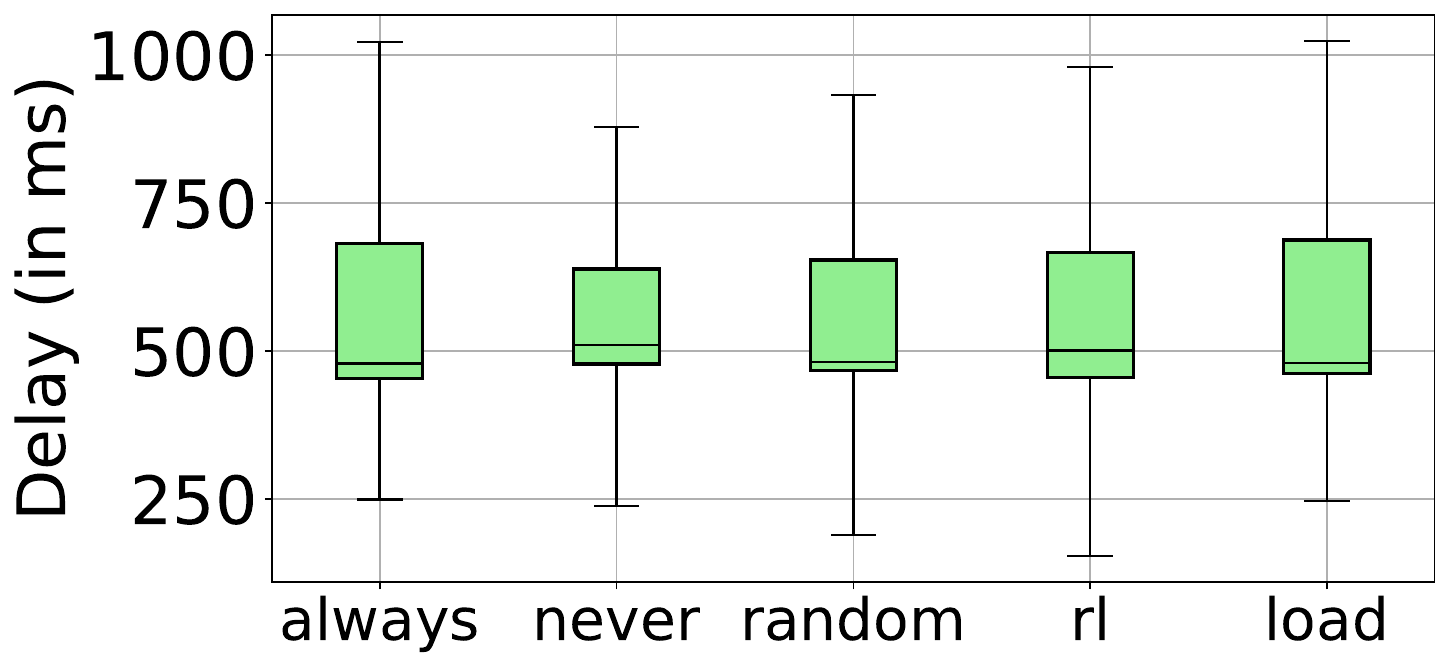}
        \caption{ORAN-xApp}
        \label{subfig:x2}
    \end{subfigure}
    \begin{subfigure}[t]{0.32\textwidth}
    \centering
        \includegraphics[scale=0.23]{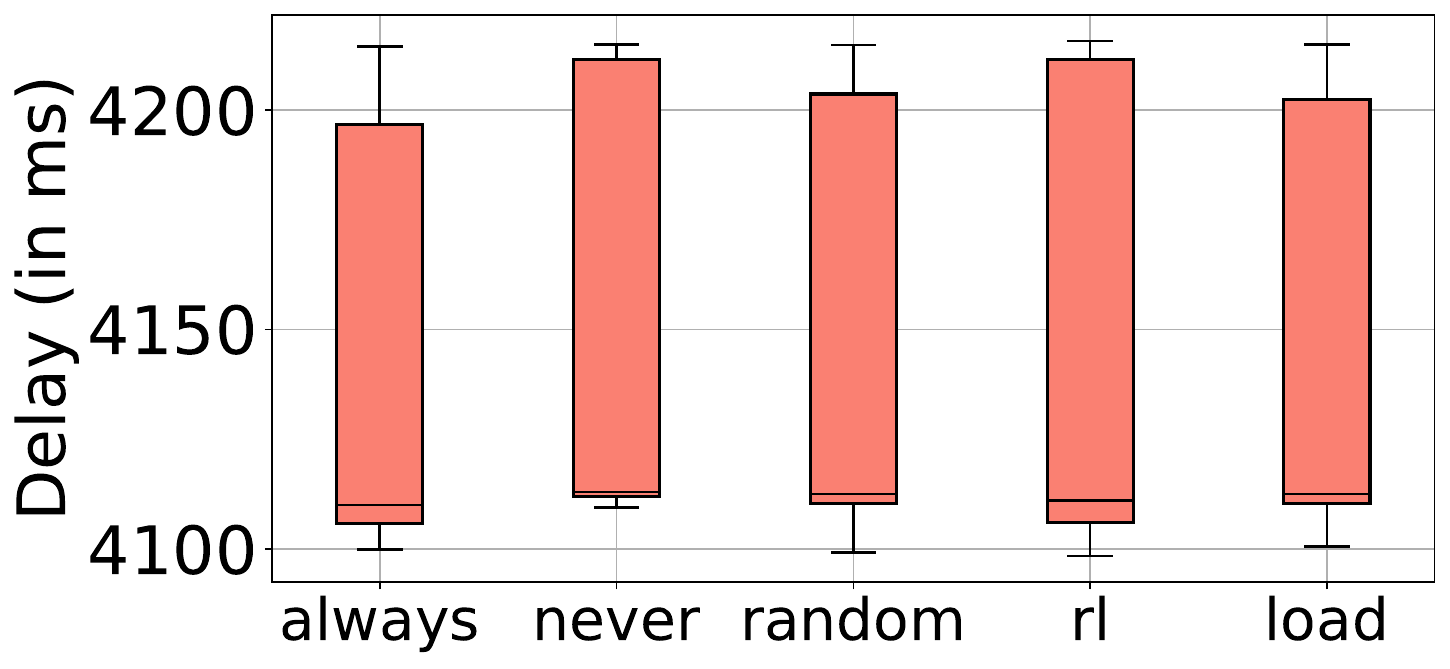}
        \caption{ORAN-rApp}
        \label{subfig:r3}
    \end{subfigure}
    \caption{Delay for different O-RAN RIC application workflows}
    \label{fig:delay_box}
\end{figure*}

We perform a numerical analysis to evaluate the efficiency and performance of different update agent implementations to decide whether to proceed with updating to the latest version of an ML model or to continue operating with the existing stable version. We employ an event-based simulator with inter-arrival and service times following a negative exponential distribution and report the results with 98\% confidence level.
A first-fit algorithm is used for resource allocation in assigning ML model instances to worker nodes.
For scaling, we adopt a monitoring-based approach, commonly used in practice. This algorithm monitors the load on each worker: if the load exceeds a predefined threshold, a new replica is scheduled for creation. On the other hand, if a replica is idle, having processed all queued requests and awaiting new ones, it is removed. If neither condition is met, no scaling action is taken.
In addition to the RL-based update agent, we also implement several baseline strategies: i) \emph{always update}, ii) \emph{never update}, iii) \emph{randomly select updates}, and iv) \emph{server load-based} update.

\emph{Always} update, \emph{never} update, and \emph{update randomly} serve as reference points to understand baseline behavior. The \emph{Server Load-based} strategy 
operates based on the current load of the node hosting the replica. When deciding whether to update an existing replica, the algorithm computes the load on the corresponding node. If this load is below a predefined threshold, the algorithm proceeds with the update; otherwise, it continues execution with the current ML model version. In the following experiments, we assume a load threshold value of 0.5. When creating new replicas, the Server Load-based strategy selects the model version at random, as the target node for deployment is not yet known and an accurate load estimate is therefore unavailable.

Our topology is similar to the studied architecture in  Figure \ref{fig:system}. In the Edge Cloud we assume 2 ORAN-worker nodes, 
and 2 ML-worker nodes for ML model inference. All of the Edge Cloud nodes are assumed to have 32 CPU cores and 32 GB of RAM.
In the Regional Cloud, Near-Real-time RIC is deployed on the ORAN site and a cloud node is allocated for ML inference, with unlimited  CPU and RAM capacities.
Same assumptions are adopted for the Central Cloud. 

\begin{table}[h]
\resizebox{\linewidth}{!}{
\begin{tabular}{|l|c|c|c|c|c|c|c|}
\hline
\rowcolor[HTML]{EFEFEF}
\textbf{Parameter}              & ML-d1 & ML-d2 & ML-x1 & ML-x2 & ML-r1 & ML-r2 \\ \hline
\rowcolor[HTML]{EFEFEF}Version                & \multicolumn{6}{c|}{0 -- 2000} \\ \hline
Avg service time (ms)           & 2--0.5     & 4--0.8     & 200--100   & 300--200   & 1000--900  & 2000--1800  \\ \hline
Avg inter-arrival time (ms)     & 3     & 4     & 350   & 525   & 1750  & 3500  \\ \hline
Spawn time                      & 3     & 3     & 100   & 100   & 1000  & 1000  \\ \hline
CPU requirement                & 1     & 2     & 16    & 8     & 32    & 32    \\ \hline
RAM requirement                & 1     & 1     & 32    & 32    & 48    & 64    \\ \hline
Disk requirement               & 0.01  & 0.02  & 0.1   & 0.2   & 1     & 2     \\ \hline
Accuracy                       & 0.7--1.0 & 0.7--1.0 & 0.75--1.0 & 0.75--1.0 & 0.8--1.0 & 0.8--1.0 \\ \hline
Stability                      & 1.0--0.7 & 1.0--0.7 & 1.0--0.7 & 1.0--0.7 & 1.0--0.7 & 1.0--0.7 \\ \hline
\end{tabular}
}
\caption{ML model parameters simulated.}
\label{table:funmodels}
\end{table}

We simulate six ML models that power O-RAN RIC functions (Table \ref{table:funmodels}). Each model is initially released as version 0.0, followed by major u and minor upgrades (1.0, 2.0, …, 10.200). 
With each update, inference accuracy increases, whereas replica stability and service time gradually degrade. 
Each major release improves inference accuracy by $\sim2\%$, but reduces replica stability by $\sim2\%$ and the service time by $\sim7\%$. 
A subversion update will improve the ML model accuracy by  $\sim 0.1\%$, to imitate a small performance update, and the stability will decrease by $\sim 0.1\%$.
For the reward function defined in Eq. (\ref{eq:reward}), the weights are set to: $w_1=0.025, w_2=1, w_3=2$. 
For Q-table update, we set $\alpha = 0.01$, $\gamma = 0.99$, the initial exploration probability $\epsilon  = 1$ , this will decay with up to a minimal $\epsilon_\text{min} = 0.001$ which will be reached after half of the scheduled events have been processed. RL settings are mainly for training and have shown good performance during hyperparameter tuning.

Fig. \ref{fig:delay_box} shows boxplots for the delays ($\mathcal{O}_1$) of the entire ORAN RIC application flows for dApp, xApp and rApp. For the ORAN-dApp in Fig. \ref{subfig:d1}, the delay for the \emph{always update } strategy is the best, as expected, because although it needs to spawn new replicas, whenever there is a new version available to update existing replicas, the new versions will have lower inference time. The \emph{never update} strategy performs worse, but is comparable to the \emph{random, rl } and \emph{load-based } update strategies. All delays result around median values of 10 to 12 milliseconds, as typical for dApp delays. 
For the ORAN-xApp in Fig. \ref{subfig:x2}, the median delay is 490 milliseconds, for the always update strategy is the lowest, together with the \emph{load-based} update strategy.  Both update strategies have the highest standard deviation, because of the added spawn time, which is higher in relation to the processing than in the dApp case.
For ORAN-rApp in Fig. \ref{subfig:r3} the delays result in a median of 4.11 seconds, where the \emph{always update} strategy results again in the lowest median, followed by the \emph{RL} strategy.

\begin{figure}
 \centering 
 \begin{subfigure}[t]{0.5\textwidth}
 \centering 
 \includegraphics[scale=0.24]{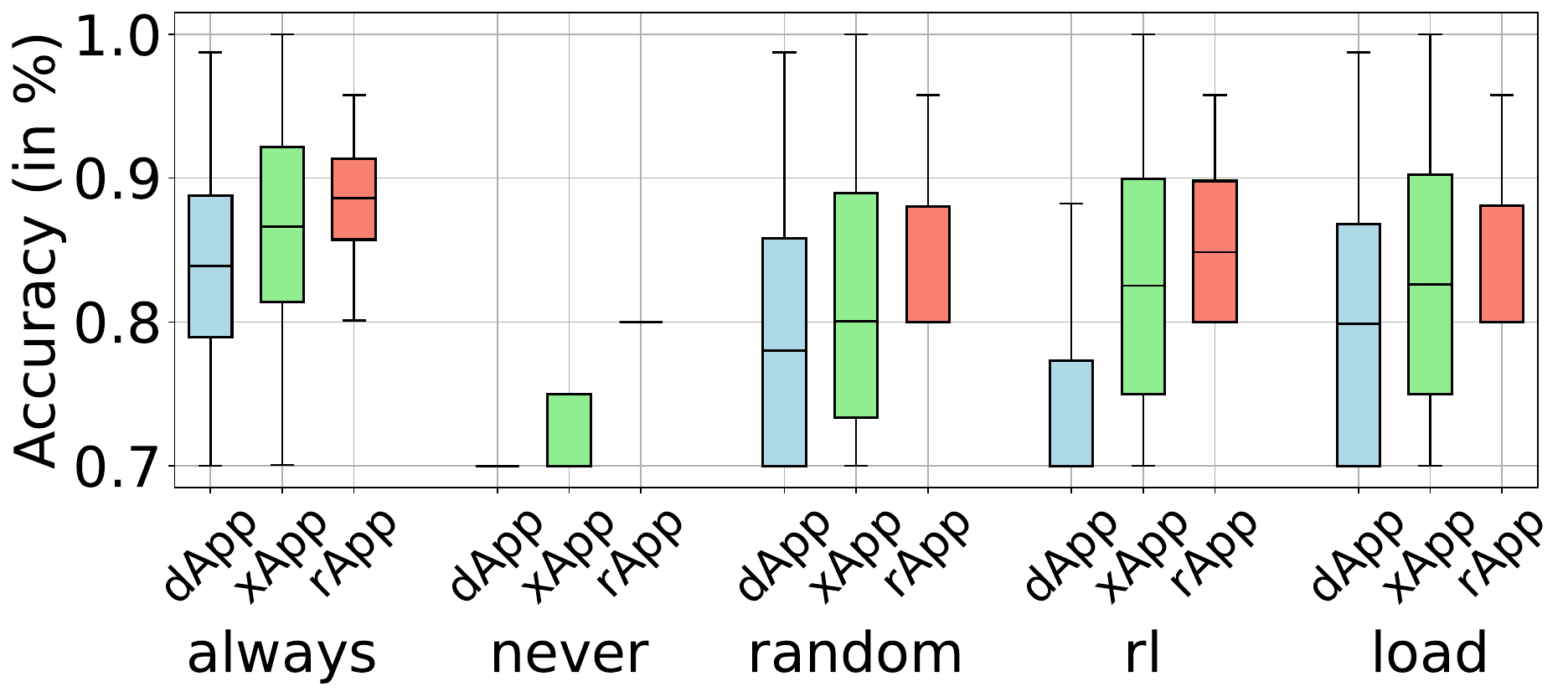}
  \caption{Accuracy  }\label{subfig:mat1}
   \end{subfigure}
   \begin{subfigure}[t]{0.5\textwidth}\centering 
 \includegraphics[scale=0.24]{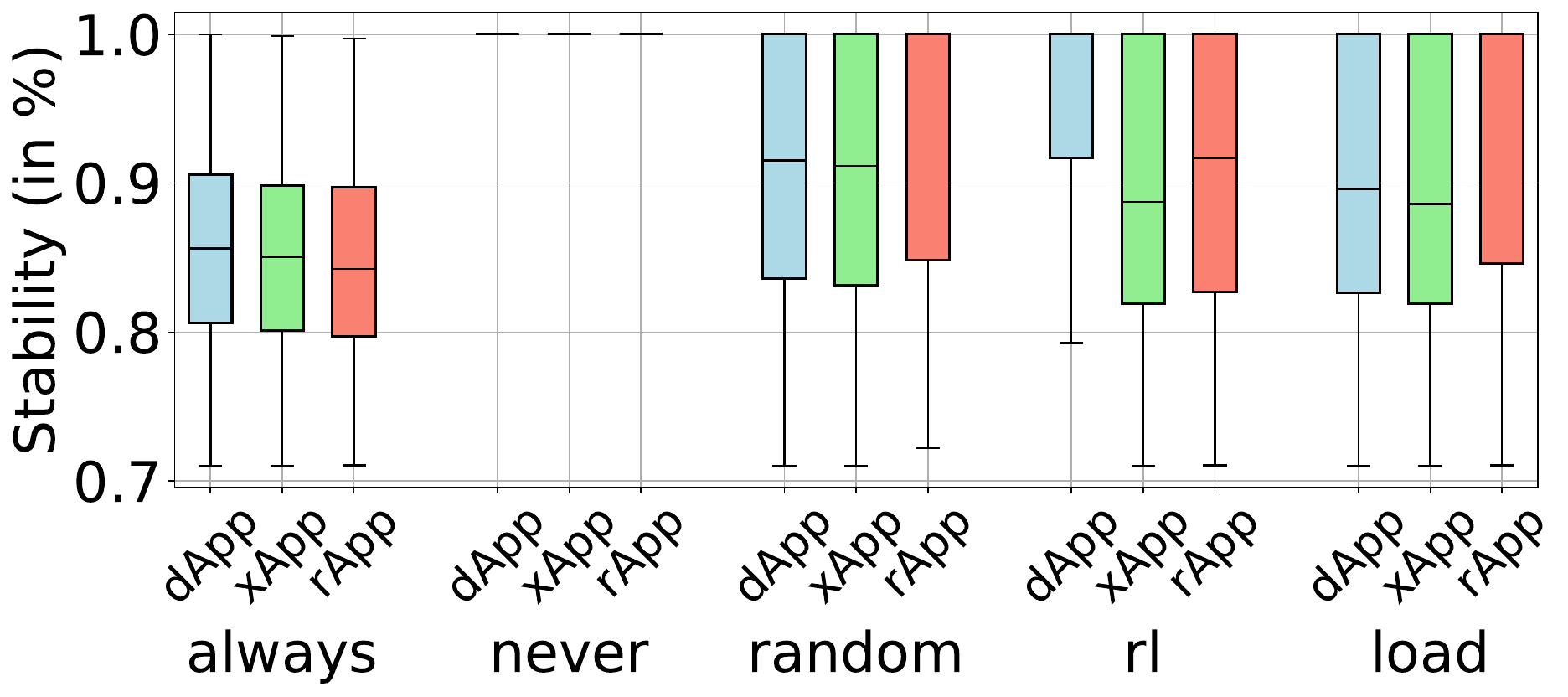}
 \caption{Stability  }\label{subfig:mat2}
   \end{subfigure}
 \caption{Performance metrics for ML Models}
\label{fig:performance_metrics_1}
\end{figure}

Fig. \ref{fig:performance_metrics_1} shows boxplots for the average accuracy ($\mathcal{O}_2$) and the average stability ($\mathcal{O}_3$) for all update strategies and all O-RAN RIC  applications.
In Fig. \ref{subfig:mat1}, the accuracy of the \emph{always update} strategy is the highest for all ORAN applications and the lowest for the \emph{never update} strategy. The \emph{load-based} strategy achieves a better median than the \emph{random update} strategy but the \emph{rl} strategy decides to prefer the stability of the system over the highest accuracy, especially for the latency constrained ORAN-dApp. For the xApp and rApp the RL agent archives better accuracy values than the other two strategies.
The stability shown in Fig. \ref{subfig:mat2} is always fixed to 1 for the \emph{never update} strategy. Always updating results in the lowest stability, but the \emph{rl} strategy achieves very high stability values for dApp and comparable stability for xApp and rApp in perspective of \emph{random update} and \emph{load-based} update strategies.

 \section{Conclusion}\label{sec:conclusion}
This paper tackled a critical gap on the path to AI-native 6G: scalable self-learning life-cycle management of the many ML models that will drive rApps, xApps, and dApps in O-RAN. We presented the first end-to-end framework that links cloud-side training with edge-side inference.  Our solution jointly optimized ML model accuracy, stability, and latency across the O-RAN control loops. Simulation results showed that RL policy learns to prioritize stability for delay-sensitive dApps while upgrading rApps/xApps at moments that minimize latency impact, thereby increasing overall network utility and reducing SLA violations compared to baselines. These findings confirm that closed-loop, telemetry-driven model versioning is essential for future SMO/RIC platforms. Future work will focus on testbed validation and security-aware reward design.




\bibliographystyle{IEEEtran}
\bibliography{mybib}

%

\end{document}